\documentclass[11pt]{iopart}
\RequirePackage[colorlinks,citecolor=blue,urlcolor=magenta,linkcolor=blue]{hyperref}
\usepackage{graphicx}
\usepackage{enumerate}
\usepackage[shortlabels]{enumitem}
\usepackage{url}
\usepackage{amsfonts}
\usepackage[thinc]{esdiff}
\usepackage{setspace}
\usepackage{verbatim}
\usepackage{dutchcal}
\usepackage{tabularx}
\usepackage{scalerel}
\usepackage{braket}
\usepackage{tikz}
\usepackage{tikz-cd}
\usetikzlibrary{svg.path}

\usepackage[normalem]{ulem}

\newcommand{\p}{\partial}

\renewcommand{\d}{\text{d}}

\def\tl{\textrm{T}}
\def\td{{\cal T}}
\newcommand{\BK}{Brown-Kucha\v{r} }
\usepackage{mathtools}
\DeclarePairedDelimiter{\abs}{\lvert}{\rvert}

\usepackage{moreenum}
\usepackage[utf8]{inputenc}

\labelformat{equation}{Eq.~(#1)} 
\labelformat{figure}{Fig.~#1} 
\labelformat{subfigure}{Fig.~\thefigure#1} 
\labelformat{table}{Tab.~#1} 
\labelformat{appendix}{Appendix #1}

\definecolor{orcidlogocol}{HTML}{A6CE39}
\tikzset{
  orcidlogo/.pic={
    \fill[orcidlogocol] svg{M256,128c0,70.7-57.3,128-128,128C57.3,256,0,198.7,0,128C0,57.3,57.3,0,128,0C198.7,0,256,57.3,256,128z};
    \fill[white] svg{M86.3,186.2H70.9V79.1h15.4v48.4V186.2z}
                 svg{M108.9,79.1h41.6c39.6,0,57,28.3,57,53.6c0,27.5-21.5,53.6-56.8,53.6h-41.8V79.1z M124.3,172.4h24.5c34.9,0,42.9-26.5,42.9-39.7c0-21.5-13.7-39.7-43.7-39.7h-23.7V172.4z}
                 svg{M88.7,56.8c0,5.5-4.5,10.1-10.1,10.1c-5.6,0-10.1-4.6-10.1-10.1c0-5.6,4.5-10.1,10.1-10.1C84.2,46.7,88.7,51.3,88.7,56.8z};
  }
}

\newcommand\orcidicon[1]{\href{https://orcid.org/#1}{\mbox{\scalerel*{
\begin{tikzpicture}[yscale=-1,transform shape]
\pic{orcidlogo};
\end{tikzpicture}
}{|}}}}

\usepackage[square,numbers,compress]{natbib}
\usepackage{lipsum}

\usepackage{bm}
\usepackage[svgnames]{xcolor}\usepackage{slashed}
\usepackage{xparse}
\usepackage{dsfont}
\usepackage[mathscr]{euscript}
\usepackage{float}
\usepackage{etoolbox}
\patchcmd{\section}
  {\centering}
  {\raggedright}
  {}
  {}
\patchcmd{\subsection}
  {\centering}
  {\raggedright}
  {}
  {}
\DeclareUnicodeCharacter{2212}{-}
\begin{document}

\title[Quantum Cosmology as a Hydrogen atom: Discrete $\Lambda$ and cyclic Universes \dots]{Quantum Cosmology as a Hydrogen atom: Discrete $\Lambda$ and cyclic Universes from Wheeler-DeWitt quantization}

\author{Dipayan Mukherjee$^1$\orcidicon{0000-0002-5267-1944}, Harkirat Singh Sahota$^2$\orcidicon{0000-0002-9389-0932} and S.~Shankaranarayanan$^3$\orcidicon{0000-0002-7666-4116}}
\address{$^1$ Raman Research Institute, C. V. Raman Avenue, Sadashivanagar, Bengaluru 560080, India}
\address{$^2$ Department of Physics, Indian Institute of Technology Delhi, Hauz Khas, New Delhi, 110016, India}
\address{$^3$ Department of Physics, Indian Institute of Technology Bombay, Mumbai 400076, India}

\ead{dipayanmkh@gmail.com}
\ead{harkirat221@gmail.com}
\ead{shanki@iitb.ac.in}

\vspace{10pt}
\begin{indented}
\item[] 27 January 2026
\end{indented}

\begin{abstract}
Building upon our recently established correspondence between quantum cosmology and the hydrogen atom~\cite{Sahota2025}, we investigate the specific sector of a negative cosmological constant ($\Lambda < 0$) in a flat FLRW universe with dust. While the positive $\Lambda$ sector~\cite{Sahota2025} yields a continuous spectrum and a single bounce, we show here that the negative $\Lambda$ sector leads to a \emph{discrete spectrum} of energy eigenvalues, effectively quantizing the cosmological constant. Within this dual description, the operator-ordering ambiguity parameter appears as the azimuthal quantum number of the hydrogen atom. A \emph{skewed Bohr correspondence} emerges for the bound states, matching classical evolution at large volumes but deviating near the bounce. By constructing wave packets from these bound states, we demonstrate that the classical Big Bang and Big Crunch singularities are resolved, and the universe oscillates between quantum bounces and classical turnaround points. The expectation values of the observables indicate a cyclic universe --- with vanishing Hubble parameter at turnarounds --- undergoing quantum bounces. This exactly solvable model offers a tractable setting to explore quantum gravitational effects in cosmology. We analyze the properties of this cyclic universe, contrasting its bound-state dynamics with the scattering states of the de Sitter case.
\end{abstract}

\section{Introduction}
Spacetime singularities in general relativity (GR), such as the \emph{Big Bang}, signify the breakdown of classical physics and underscore the need for a quantum theory of gravity~\cite{Penrose1965,Hawking:1971vc}. While a complete theory remains elusive, symmetry-reduced cosmological models offer a tractable avenue to explore fundamental quantum effects like singularity resolution \cite{Vilenkin:1982de,Vilenkin_ordering,Rubakov:1984bh,Coule:2004qf,kiefer_classical_2006,Pedram:2007fm,Pedram:2007ud,kreienbuehl_singularity_2009,BarberoG:2010oga,Husain:2011tm,Kiefer2012-dq,Kamenshchik:2013naa,bergeron_singularity_2015,kiefer_singularity_2019,kiefer_singularity_2019-1,Malkiewicz:2019azw,Husain:2019nym,Vieira:2020azi,Sahota_Infrared,Gielen:2022dhg,Alexandre:2022ijm,Alexandre:2022npo,Vaz:2022jzq,Vaz:2022kdq,Vaz:2022rxv,Thebault:2022dmv,Mukherjee:2023qan,Sahota2023,Gielen_PRL,Piazza:2025fbt}. The canonical quantization of a cosmological model leads to the Wheeler-DeWitt (WdW) equation \cite{wheeler_superspace_nodate,dewitt_quantum_1967}, and a relational approach is typically adopted to arrive at the notion of time in the quantum theory \cite{Tambornino:2011vg,Hoehn:2019fsy,Chataignier:2021ncn}.

In this work, we investigate the canonical quantization of a spatially flat Friedmann-Lemaître-Robertson-Walker (FLRW) universe containing pressureless dust (via the \BK formalism \cite{brown_dust_1995}) and a cosmological constant $\Lambda$, treated dynamically within the unimodular gravity framework \cite{Henneaux:1989zc,Unruh:1988in,Smolin:2009ti}. Strikingly, WdW quantization \cite{wheeler_superspace_nodate,dewitt_quantum_1967} of this system, when unimodular time is employed as the clock, is exactly mapped to the radial dynamics of a non-relativistic hydrogen atom \cite{Sahota2025}. In this correspondence, the volume of the universe maps to the radial coordinate and $\Lambda$ assumes the role of the atom's energy. The search for exact solutions in quantum cosmology is vital for understanding of singularity resolution, and this mapping allows us to import the analytical tools of atomic physics into the investigations of quantum cosmology. For earlier investigations into similar scenarios, see \cite{Amemiya_2009,Maeda:2015fna,Ali:2018vmt}, where the Brown-Kucha\v{r} dust is used as relational clock, while \cite{Gryb:2018whn,Gielen:2020abd,Gielen:2021igw,Gielen:2022tzi} consider unimodular clock for FLRW universe with scalar field and cosmological constant.

This correspondence between the WdW equation for a dust-filled FLRW universe and the radial Schr\"odinger equation of the non-relativistic hydrogen atom was first introduced in our previous work~\cite{Sahota2025}, where the analysis focused primarily on the case of a positive cosmological constant ($\Lambda > 0$). In the language of the hydrogen correspondence, this sector maps to the \emph{scattering states} of the atom. Physically, this describes a universe with a continuous energy spectrum that undergoes a single quantum bounce before expanding indefinitely, while it asymptotically recovers the classical behavior away from the bounce.

However, the hydrogen atom formalism suggests the existence of a distinct and physically richer sector: the \emph{bound states}, corresponding to a negative cosmological constant ($\Lambda < 0$). This regime was not analyzed in detail in Ref.~\cite{Sahota2025} and offers fundamentally different quantum cosmological implications.

Moreover, the operator ordering ambiguity remains a critical issue in canonical quantum gravity, as different orderings generally lead to inequivalent physical scenarios \cite{Kontoleon:1998pw,Steigl:2005fk,Sahota:2023kox,Mondal:2025qyd}.
We address this issue in this work by incorporating \emph{general Hamiltonian operator ordering} characterized by an ambiguity parameters, and analyze its role in the quantum dynamics. To this end, we extend the quantum cosmology--hydrogen atom correspondence by keeping the operator ordering general, which in turn enables us to identify the emergence of ambiguity in the energy spectrum. Interestingly, we find that operator-ordering ambiguity parameter maps to the azimuthal quantum number of the hydrogen atom in the dual description.

We then perform a dynamical analysis of this $\Lambda < 0$ sector. We show that unlike the scattering case, this regime leads to:
\begin{enumerate}
    \item  \emph{A Discrete Spectrum}: The cosmological constant $\Lambda$ cannot take arbitrary values but is quantized.
    \item \emph{Cyclic Evolution:} By constructing wave packets from these bound states, we demonstrate that the universe undergoes cycles of bounces and turnarounds, avoiding Big Bang and Big Crunch singularities.
    \item \emph{Bohr Correspondence}: In the limit of large quantum numbers (corresponding to a large maximum volume), a quantum-classical correspondence emerges, though it is modified near the bounce, a feature we term a \emph{skewed Bohr correspondence}.
\end{enumerate}

The negative $\Lambda$ sector holds significant theoretical interest due to its connections to holography~\cite{Maldacena:1997re,Klebanov:1999tb,Skenderis:1999nb} and quantum gravity toy models \cite{solodukhin2016metric,Meert:2025cct}. Moreover, the system under consideration can describe the interior of an AdS-Schwarzchild black hole \cite{Tibrewala:2007ex,Vaz:2008jm} {\it \'a la} AdS generalization of Oppenheimer-Snyder collapse \cite{kiefer_singularity_2019,PhysRevD.101.026016,PhysRevD.103.064074,Piechocki:2020bfo,Khodabakhshi:2025fmf}.  
While observations favor $\Lambda > 0$, AdS cosmologies ($\Lambda < 0$) remain critical for theoretical studies, especially given recent JWST observations~\cite{Menci_2024, Adil_2023, Wang_2025,Wang:2025dtk}. The study of AdS quantum cosmology is sparse in the literature, a few quantum cosmology analyses for negative $\Lambda$ universe in related scenarios include~\cite{Ali:2018vmt}, where the dust variable served as a clock in AdS-dust quantum cosmology and~\cite{Gryb:2018whn}, where the authors consider unimodular gravity with a scalar field. Classically, the universe evolves from Big Bang to Big Crunch, but this hydrogen-like correspondence yields exact quantum solutions, leading to a quantized $\Lambda$ spectrum and singularity free quantum universe, offering a rare analytical window into quantum gravity.

This paper is organized as follows. In Sec.~\ref{sec:2}, we summarize the hydrogen atom correspondence of an FLRW universe with dust and $\Lambda$, introduced in Ref.~\cite{Sahota2025}. We then extend the correspondence by incorporating general operator ordering in the quantum Hamiltonian. In Sec.~\ref{sec:3}, we demonstrate the resolutions of the classical Big Bang and Big Crunch singularities in light of Bohr's correspondence for the bound states. In Sec.~\ref{sec:4} we construct coherent wave packets from the bound states and study the quantum evolutions of relevant  cosmological quantities. We end with a summary and conclusions in Sec.~\ref{sec:5}.

\section{Hydrogen-atom correspondence}
\label{sec:2}
The total action comprising of the Einstein-Hilbert term, \BK dust, and unimodular action for the cosmological constant $\Lambda$ (in $\hbar = c = 8 \pi G =1$ units) is:
\begin{align}
\label{eq:action}
S =& \frac{1}{2} \int \d^4 x \sqrt{-g} \left( \mathcal{R} - 2 \Lambda \right)+ \int \d^4 x \Lambda\p_\alpha \tl^\alpha-\frac{1}{2} \int \d^4 x \sqrt{-g} ~ \rho \left( g^{\alpha \beta}u_\alpha u_\beta + 1 \right),
\end{align}
where $\mathcal{R}$ is the Ricci scalar, $\rho$ and $u^\alpha$ are the energy density and four-velocity of the \BK dust \cite{brown_dust_1995}, and $\tl^\alpha(x)$ is a dynamical vector field associated with the unimodular sector. For the FLRW metric, $\d s^2 = - N^2(t) \d t^2 + a^2(t) \delta_{ij} \d x^i \d x^j$, only the time components of $u^0 = -\dot{\td}$ and $\tl^0 = \dot{\tl}$ (where we redefine the scalar field as $\tl(t)$) contribute, with $\td(t)$ and $\tl(t)$ being the dynamical variables for the dust and unimodular sectors, respectively. Our analysis employs the phase-space variables $\{v \equiv a^3, \tl, \td, N; P_v, P_{\tl}, P_{\td}, P_N\}$, where $P_i$ are the conjugate momenta. 

The action \ref{eq:action} leads to the classical Hamiltonian constraint, or  equivalently the first Friedmann equation:
\begin{align}
 H = N\left[ - ({3}/4) \, vP_v^2 + \rho_0  + vP_{\tl}  \right] \approx 0,
\end{align}
where we set the spatial volume to unity, $\rho_0 = P_{\td}$ is a constant of motion representing the dust's total comoving energy density, and the physical dust energy density is $\rho = \rho_0 v^{-1}$. In the quantum theory, $\tl$ emerges as a suitable reference clock, facilitating a unitary evolution of the system \cite{Sahota2025}. Adopting $\tl$ as the clock, a convenient lapse choice is $N=v^{-1}$, which establishes a linear relationship between $\tl$ and the coordinate time $t$ (i.e., $\tl = t + \text{constant}$).
With this choice, the Hamiltonian constraint leads to WdW equation:
\begin{align}
  \hat{H}_u\Psi(v,\tl) &= i \, \frac{\partial\Psi(v,\tl)}{\partial \tl} \, , \label{HuEV}
\end{align}
where the unimodular Hamiltonian $\hat{H}_u$ is an quantum operator corresponding to the classical quantity $-(3/4) P_v^2 + \rho_0 v^{-1}$.
The kinetic term of gravitational Hamiltonian often involves products of configuration variables and their momenta, leading to operator ordering ambiguities~\cite{Sahota2023,Kontoleon:1998pw,Steigl:2005fk,Sahota:2023kox,Mondal:2025qyd}. To systematically address this, we adopt a generalized ordering prescription introduced in \cite{kiefer_singularity_2019}, leading to:
\begin{align}
  \hat{H}_u \equiv \frac{3}{4}v^{p+q}\frac{\partial}{\partial v}v^{-p}\frac{\partial}{\partial v}v^{-q} + v^{-1}\rho_0 \label{Hu}
\end{align}
where the ambiguity is parameterized by the real positive numbers $p$ and $q$, and Hamiltonian operator is Hermitian on $L^2(\mathbb{R}^+,v^{-p-2q} \d v)$. Substituting the ansatz $\Psi(v,\tl) = e^{i\Lambda\tl}\psi_\Lambda(v)$ into \ref{HuEV} with $\hat{H}_u$ from \ref{Hu}, and performing a redefinition $\psi_\Lambda(v) = v^{p/2 + q} \chi(v)$, the stationary WdW equation becomes:
\begin{align}
    \frac{\d^2\chi(v)}{\d v^2} + \left[ - \frac{p(2 + p)}{4 v^2}  
    + \frac{4}{3} \left( \frac{\rho_0}{v} + \Lambda \right) \right] \chi(v) = 0.\label{WdW}
\end{align}
Remarkably, this equation exhibits an exact mathematical correspondence with the radial Schr\"odinger equation of a non-relativistic hydrogen atom \cite{Sahota2025,landau_quantum}. The mapping is as follows: the universe's volume $v$ corresponds to the radial coordinate $r$; the cosmological constant $\Lambda$ is analogous to the energy $E$ (specifically, $4\Lambda/3 \leftrightarrow 2 m_e E/\hbar^2$); the dust parameter $\rho_0$ relates to the strength of the Coulomb interaction (effectively, $4\rho_0/(3v)$ plays the role of an attractive Coulomb potential $-Ze^2/r$ if $\rho_0 > 0$, implying $4\rho_0/3 \leftrightarrow Z e^2 (2m_e/\hbar^2)$); and crucially, the operator ordering parameter $p$ corresponds to twice the azimuthal quantum number, $p \leftrightarrow 2\ell$. Note that, for a fluid with equation of state $\omega$, the potential term in the WdW equation will be of the form $\rho_0v^{-(1+\omega)}$, which will break the hydrogen atom correspondence.

This powerful analogy extends directly to the solutions. For a positive cosmological constant ($\Lambda>0$), the normalized stationary states are given by:
{\small 
\begin{eqnarray}
\Psi_{k}(v) &=&
\sqrt{\frac{2^{1+p}}{\pi}} k^{1+\frac{p}{2}}
     \e^{\frac{\pi\rho_0}{3k}}
     \frac{\big|\Gamma\left(1+\frac{p}{2} - \frac{2i\rho_0}{3k}\right)\big|}
     {\Gamma(p+2)}v^{1+p+q}
     \nonumber\\
    & & \times
     \e^{-ikv} {}_1F_1\left(1+\frac{p}{2}+i\frac{2\rho_0}{3k},p+2,2ikv\right),
     \label{GenpPosSS}
\end{eqnarray}
}
where $k \equiv 2\sqrt{\Lambda/3}$ and ${}_1F_1$ is the confluent hypergeometric function. These solutions are analogous to the scattering states of hydrogen atom ($E>0$), implying that $\Lambda$ possesses a continuous spectrum for $\Lambda>0$. (Details can be found in~\cite{Sahota2025}.) Cosmologically, these states represent ever-expanding (or contracting, depending on boundary conditions) universes permissible for any positive $\Lambda$. This inherent stability of states for any $\Lambda>0$ within this model contrasts with discussions of de Sitter instabilities in other models~\cite{Moreau:2018lmz}, suggesting that, within minisuperspace approximation, a positive $\Lambda$ \emph{does not} inherently lead to decay.

For a negative cosmological constant ($\Lambda < 0$), the normalized stationary states are:
{\small 
\begin{align}
\Psi_{n}(v) =&
\sqrt{\frac{(n-1-\ell)!}{2n(n+\ell)!}} 
\left(\frac{4\rho_0}{3n}\right)^{\ell+\frac{3}{2}} 
v^{2\ell+1+q} 
     e^{-2\rho_0v/3n}
     L_{n-1-\ell}^{2\ell+1}\left(\frac{4\rho_0v}{3n}\right), 
     \label{GenpNegSS}
\end{align}
}
$n$ plays the role of the principle quantum number of the hydrogen atom, 
$ L_{n-1-\ell}^{2\ell+1}$ is a generalized Laguerre polynomial of degree $n-1-\ell$,
and the ambiguity parameter $p$ maps to the azimuthal quantum number, $\ell \leftrightarrow p/2$. The other ordering parameter $q$ leaves no imprint on the kinematical structure of the quantum model, and is not expected to appear in quantum dynamics as well \cite{Mondal:2025qyd}. These solutions map to the bound states of a hydrogen atom ($E<0$), leading to a discrete spectrum for $\Lambda$:
\begin{align}
    \Lambda_n = -\frac{\rho_0^2}{3n^2},\quad
    n \in \mathbb{Z}_+ 
    \&\ 1 + \frac{p}{2}<n\ 
     \&\ \frac{p}{2} \in  \mathbb{Z}_+. 
\end{align}
This quantization of $\Lambda$ is a direct quantum gravitational prediction --- only specific negative values of the cosmological constant are allowed for these \emph{bound-state universes}. This is an important cosmological implication, suggesting that if the universe were to possess a negative $\Lambda$ and be described by such a state, its fundamental vacuum energy (or curvature scale) would be intrinsically discrete, determined by the dust content ($\rho_0$) and quantum numbers $(n, \ell)$. This resonates with findings of discrete spectra in other quantum gravity contexts, like planar AdS black holes \cite{Gielen_PRL}. The principal quantum number $n$ dictates the \emph{energy level} $\Lambda_n$. The operator ordering parameter $p=2\ell$ (azimuthal quantum number analog) constrains the allowed $n$ values ($n>\ell+1$) but, interestingly, does not shift the $\Lambda_n$ levels themselves, reminiscent of the degeneracy in the hydrogen atom. However, unlike atomic degeneracy, each choice of $p$ (operator ordering) defines a distinct quantum theory, which must be fixed apriori or constrained empirically. The structure of the quantum model itself imposes conditions: $p$ must be such that $\ell=p/2$ is a non-negative integer or half-integer for $\Lambda<0$ to ensure normalizability of Laguerre polynomials, while it can be a positive real for $\Lambda>0$. The second ordering parameter, $q$, does not influence the kinematical structure or the spectrum \cite{Mondal:2025qyd}, simplifying the ambiguity space. The exact solvability provided by this analogy makes this model a valuable laboratory for studying foundational issues in quantum cosmology, offering clear interpretations for quantum states of the universe.

In this letter, we focus on the physical implications of the discrete $\Lambda < 0$ spectrum, including the implications near space-time singularity. The complementary case of a continuous $\Lambda > 0$ spectrum, which admits scattering-like universe states, is treated comprehensively in~\cite{Sahota2025}.

\section{Singularity avoidance}
\label{sec:3}
\begin{figure}
  \centering
  \includegraphics[width=0.5\columnwidth]{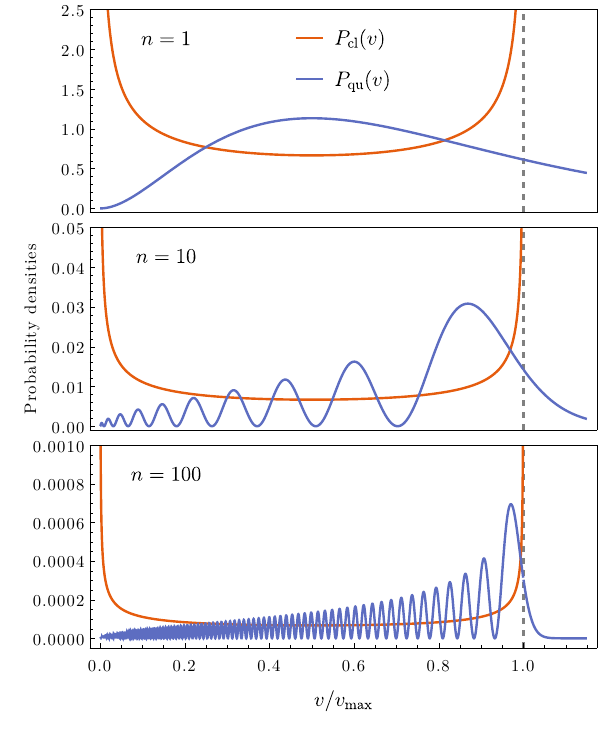}
  \caption{Comparison of $P_{\text{qu}}(v)$ (blue) and $ P_{\text{cl}}(v)$ (red) of the volume parameter for different quantum numbers.}
  \label{fig:PDFs}
\end{figure}

While the quantized universe with a discrete spectrum for $\Lambda_n$ appears distinct from its classical counterpart, we find that a classical description does emerge in the appropriate limit. To demonstrate this, we first establish the classical benchmark. For a given negative cosmological constant $\Lambda = -|\Lambda|$, the classical solution to the Friedmann equation describes a recollapsing universe. The evolution of its volume, $v$, can be parameterized as:
\begin{align}
v(\tau)
  &=  v_{\text{max}} \sin^2 \left({{3 \abs{\Lambda}}}\tau/2 \right),\\
  \tl(\tau)
  &=  v_{\text{max}}\tau/2  
    - \tl_{\text{max}} \sin\left( \sqrt{ 3\abs{\Lambda}} \tau \right),
\end{align}
where the parameter $\tau$ is in the domain $0 \leq \tau \leq2 \pi/\sqrt{3 \abs{\Lambda}}$, $v_{\text{max}} \equiv \rho_0/|\Lambda|$ is the maximum volume at $\tl_{\text{max}} \equiv \pi v_{\text{max}}/({2 \sqrt{3} |\Lambda|^{1/2}})$. This classical universe begins at a Big Bang singularity ($v=0$), expands to $v_{\text{max}}$, and eventually recollapses to a Big Crunch singularity at $\tl = 2 \tl_{\text{max}}$. The classical probability density of finding the universe at a given volume $v$, denoted $P_{\text{cl}}(v)$, is proportional to the time it spends there. This yields the normalized U-shaped distribution \cite{Robinett_1995,Robinett_2002}
\begin{align}
 P_{\text{cl}}(v) = \frac{1}{\pi\sqrt{v(v_\text{max} - v)}},
\end{align}
which diverges at the classical turning points $v=0$ and $v=v_{\text{max}}$, reflecting the slow rate of change of volume at these points.
We compare $P_{\text{cl}}(v)$  with the quantum probability density for \emph{stationary states}, defined as 
\begin{align}
P_{\text{qu}}(v) = v^{-p -2q}|\Psi_n(v)|^2 .    
\end{align}
This respects the appropriate measure on the configuration space. For simplicity, we present results for the ordering $p=0$ ($\ell=0$), as states with $p>0$ show similar large-$v$ behavior. The comparison, shown in \ref{fig:PDFs} for large quantum numbers $n$, is revealing. For large $n$, the oscillatory quantum probability distribution envelops the classical one, with its peaks closely tracking the U-shaped classical curve away from the origin. The amplitude of the quantum oscillations grows with volume, with the final and largest peak occurring just inside the classical turning point $v_{\text{max}}$. Beyond this point, $P_{\text{qu}}(v)$ decays exponentially into the classically forbidden region.

This comparison has important consequences for our understanding of the quantum universe. First, \emph{the classical singularity is resolved}. For all quantum numbers $n$, the quantum probability density vanishes at the origin ($P_{\text{qu}}(v \to 0) \to 0$), in stark contrast to the diverging classical probability. In accordance with the DeWitt criterion \cite{dewitt_quantum_1967}, the universe has zero probability of existing at zero volume. This quantum avoidance of the Big Bang and Big Crunch singularities is the most significant departure from the classical picture and enables a coherent description of a cyclic universe, where each classical singularity is replaced by a quantum bounce.

Second, the quantum universe exhibits non-classical behaviors even at its largest size. The non-zero probability density beyond $v_{\text{max}}$ signifies that the universe can \emph{quantum tunnel} to a volume larger than classically permitted. This effect, though suppressed, alters the nature of the classical turning point.
Lastly, these results refine our understanding of the quantum-classical transition. The model adheres to \emph{``skewed" Bohr correspondence principle}: correspondence to classical physics is achieved at large volumes (for large $n$), but it fundamentally breaks down at small volumes where quantum gravity effects dominate and resolve the singularity. This \emph{skewing} is not a failure of correspondence but rather a direct physical signature of how the quantum nature of spacetime corrects GR pathologies.

\section{Evolution of observables}
\label{sec:4}
The above stationary-state analysis suggests that classical singularities are resolved and that a classical description emerges for large quantum numbers. To investigate this further, we construct wave packets by superposing the stationary states $\Psi_n(v)$:
\begin{align}
\label{eq:Superposition}
    \Psi(v,\tl)=\sum_n  C_n \, \Psi_n(v)e^{i\Lambda_n \tl},
\end{align}
where the coefficients $C_n$ form a distribution peaked at a large principal quantum number $\bar{n}$.\footnote{As the ordering ambiguity restricts the allowed lower energy levels as $n>p/2+1$, it is not expected to leave any imprints on the semiclassical evolution.} To probe the system's semiclassical behavior, we employ generalized coherent states, specifically Klauder-Crawford (KC) states \cite{Klauder:1995yr,Bellomo:1998zq,Fox_Coherent,Crawford_2000,Chandran:2018wwc,Bateman,Mavromatis,Gur2005-dr}. The distribution function for these states are:
\begin{align}
\label{def:KC}
C^{\mathrm{KC}}_n= \frac{\mathrm{ N}(s,\alpha) 
    \, {\alpha \, s^{n-1}}}{\Gamma\left(\frac{n+1}{\alpha}\right)} \, ,
\end{align}
where $\mathrm{N}(s,\alpha)$ is normalization coefficient, and $s, \alpha$ are parameters of the distribution. The width of the distribution is related to the mean by $\Delta n=\sqrt{\alpha\bar n}$, hence small width can be achieved by small $\alpha$ for the same mean. KC states are well-suited for preserving classical-like trajectories in the quantum evolution~\cite{Crawford_2000} as these states resolves identity operator and temporally stable~\cite{Klauder:1995yr}. 
\begin{figure}
  \centering
  \includegraphics[width=0.6\columnwidth]{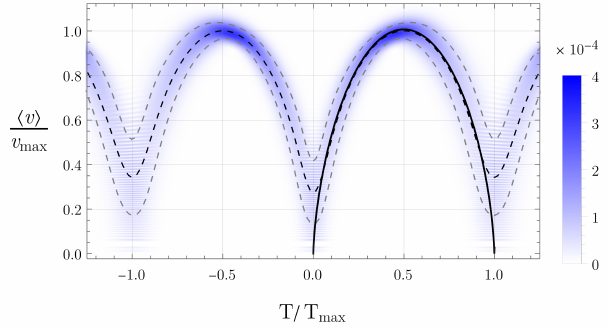}
  \caption{Quantum dynamics of AdS-dust universe for the KC wave states with $s=10^{64},$ $\alpha=1/36$ that corresponds to $\bar{n}=100$. $\rho_0=1$.
  Dashed black line represents expectation value of volume variable with dashed gray lines representing quantum uncertainty $(\braket{v}\pm\Delta v)$ while solid line depicts classical behavior. The heatmap shows the probability distribution associated with the wave packet.}
  \label{fig:Expv}
\end{figure}

The resulting quantum dynamics, depicted in \ref{fig:Expv}, reveal a universe that is both singularity-free and cyclic. The wave packet evolution shows that the classical Big Bang and Big Crunch singularities are replaced by smooth \emph{quantum bounces}. The universe does not collapse to zero volume; instead, the expectation value of the volume $\braket{v}$ turns around at a finite minimum value, and the wave packet remains well-behaved and non-singular at all times. This dynamic resolution of the singularity is a hallmark prediction of many quantum cosmology models, most notably \emph{Loop Quantum Cosmology (LQC)} where a similar \emph{Big Bounce} emerges from the underlying discrete quantum geometry~\cite{Bojowald:2001xe,Ashtekar:2006wn,Ijjas:2019pyf}. The emergence of a robust bounce in our continuous WdW framework provides a valuable point of comparison.

Following each bounce, the universe enters an expanding phase that eventually reaches a maximum volume before recollapsing, precisely mirroring the classical trajectory at large volumes, as seen in \ref{fig:Expv}. During this semiclassical phase, the wave packet is sharply peaked around the classical path, with minimal dispersion. However, as the universe approaches a bounce, quantum effects become dominant: the wave packet disperses, reflecting large quantum fluctuations, and exhibits ``ringing'' --- rapid oscillations in its probability distribution. This ringing is a characteristic interference pattern between the collapsing and expanding branches of the universe's wave function. For KC states \ref{def:KC}, this ringing is primarily confined to the bounce regions, allowing the universe to regain its classical character at large volumes.

The key outcome of this dynamical analysis is that \emph{the quantization of the AdS-dust universe leads to a perpetually cyclic cosmology}. The classical trajectory, which exists for a finite duration between two singularities, corresponds to just a single cycle of this eternal quantum evolution. This finding establishes a compelling parallel with other areas where quantum gravity effects near singularities are studied. For instance, a strikingly similar cyclic dynamic, driven by a quantum bounce that resolves a classical singularity, has been observed in the quantum description of the interior of planar AdS black holes~\cite{Gielen_PRL}. Similar coherent cyclic evolutions are also observed for AdS universes in LQC~\cite{Bojowald:2004kt,Bojowald:2007abc,Bentivegna:2008bg}. The appearance of such cyclic behavior in disparate systems suggests it may be a more general feature of how quantum gravity regulates classical pathologies.

\begin{figure}
  \centering
  \includegraphics[width=0.55\columnwidth]{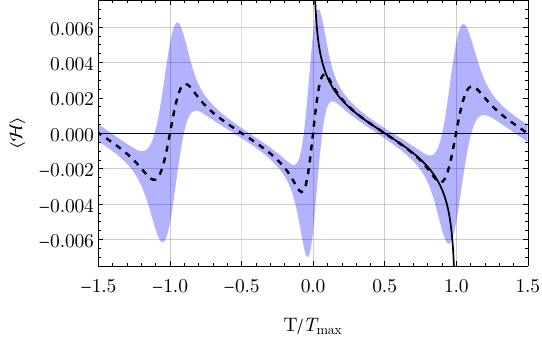}
  \caption{Expectation value of the Hubble parameter operator $\hat{\mathcal{H}}=-\hat{p}_v/2$ (dashed line) for KC states, with uncertainty $\braket{\mathcal{H}}\pm\Delta \mathcal{H}$ indicated by the shaded region and the classical Hubble parameter (solid curve), for parameters as in \ref{fig:Expv}.}
  \label{fig:Hubble}
\end{figure}

To connect our model's dynamics to observable quantities, we compute the expectation value of the effective Hubble parameter $\braket{\hat{\cal H}}$. The Hubble parameter geometrizes the universe's expansion, and its quantum behavior provides a sharp contrast to the classical picture. \ref{fig:Hubble} illustrates the evolution of $\braket{\hat{\cal H}}$ for the Klauder-Crawford wave packet (with parameters $\rho_0=1$, $s=10^{64}$, $\alpha=1/36$ that leads to $\bar{n}=100$ and $\Delta n=1.5$) and compares it to its classical counterpart.

Classically, the Hubble parameter diverges at both the Big Bang and Big Crunch singularities, signaling a catastrophic breakdown of the theory. In the quantum evolution, however, a crucial difference emerges --- as the universe approaches the point of a classical singularity, the expectation value of the Hubble parameter goes to zero, $\braket{\hat{\cal H}} \to 0$, before changing sign. This behavior is the definitive signature of a \emph{quantum bounce}. The infinite expansion and contraction rates of the classical singularities are replaced by a smooth, finite turnaround. Away from this bounce, the quantum-corrected Hubble parameter's evolution rapidly converges to the classical trajectory, reinforcing the model's adherence to quantum-classical correspondence in the appropriate regime.

\section{Conclusions}
\label{sec:5}
We have presented an exactly solvable model of quantum cosmology that exhibits a cyclic evolution for a negative cosmological constant. This work complements the analysis presented in Ref.~\cite{Sahota2025} by focusing on the bound-state sector of the `cosmological hydrogen atom'. While the formalism is shared, the physical predictions for $\Lambda < 0$ are qualitatively different from the $\Lambda > 0$ case. We have shown that the anti-de Sitter case turn out to be analogous to bound states in hydrogen atom, and the cosmological constant is quantized as $\Lambda_n = -\rho_0^2/3n^2$.
This is a key prediction where vacuum energy becomes discrete due to quantum gravitational effects. Unlike string-theoretic discreteness from flux compactifications~\cite{Bousso:2000xa},  this spectrum arises directly from the self-adjointness of quantum observables, offering a new mechanism for $\Lambda$-discreteness.

We note that other exactly solvable frameworks exist in the literature. For instance, models involving scalar fields \cite{Gryb:2018whn,Gielen:2020abd,Gielen:2021igw} or specific mappings where the roles of dust and $\Lambda$ are swapped \cite{Amemiya_2009,Maeda:2015fna} can lead to a \emph{harmonic oscillator} analogy~\cite{Ali:2018vmt}.
However, the distinguishing feature of the present work is the rigorous mapping to the \emph{Hydrogen Atom} (Coulomb potential), rather than the Harmonic Oscillator. This is non-trivial, as it leads to a qualitatively different quantization condition for the cosmological constant in the scalar field models, while the latter cases corresponds to the quantization of the dust energy. While an oscillator mapping implies an equidistant spectrum for the dust energy, our hydrogenic mapping predicts a \emph{Rydberg-like spectrum} ($\Lambda_n \propto -1/n^2$) for the vacuum energy. This provides a novel spectral profile for the cosmological constant itself that is distinct from previous scalar field or oscillator-based frameworks.

The dynamics of the model reveal a non-singular, cyclic quantum universe.  By analyzing wave packets constructed from the analytical solutions, we have shown that the classical Big Bang and Big Crunch singularities are replaced by quantum bounces. The singularity resolution leads to a perpetually cyclic cosmology, where the finite lifespan of the classical universe represents but a single cycle. This provides an exactly solvable cyclic bouncing cosmology in the WdW framework, offering a stringent benchmark for comparisons to LQC~\cite{Bojowald:2004kt,Bojowald:2007abc,Bentivegna:2008bg} and other quantum gravity approaches. 

The powerful solvability of this \emph{hydrogen atom of cosmology} provides a clear theoretical laboratory for investigating foundational questions. 
Crucially, the hydrogen analogy maps operator-ordering ambiguities to the azimuthal quantum number, offering fresh constraints for canonical quantum gravity \cite{Bojowald:2014ija,Giesel_2006,MenaMarugan:2011me}. Particularly, the ordering choice affects the lowest energy level according to $n>p/2+1$. Since large $n$ states are crucial to recover the semiclassical behavior --- as seen from the skewed Bohr correspondence --- the ordering ambiguity is not expected to affect the semiclassical cyclic evolution of the universe.
Though the focus of the current work is on $\Lambda < 0$, the framework’s tractability enables future studies of observable signatures (e.g., quantum fluctuations in the $\Lambda > 0$ continuum~\cite{Sahota2025}). The present system can further model the interior of an AdS-Schwarzchild black hole \cite{Tibrewala:2007ex,Vaz:2008jm} that describes the dynamics of dust shell \cite{kiefer_singularity_2019,PhysRevD.101.026016,PhysRevD.103.064074,Piechocki:2020bfo,Khodabakhshi:2025fmf}. Our work also suggests that analogous, solvable models~\cite{Chandran:2025azu} could be instrumental in search for observational signatures of quantum gravity.

It is worth noting that while pressureless dust is not the dominant energy component in the very early universe (where radiation dominates), its inclusion here is mathematically essential to the hydrogen atom analogy. The dust energy density scales as $\sim v^{-1}$, which provides the necessary Coulomb-like potential in the Hamiltonian. A radiation component would introduce a potential scaling as $\sim v^{-4/3}$, which would break the exact solvability of the hydrogenic spectrum. Therefore, the current model should be viewed as a foundational, exactly solvable `toy model'—analogous to the Hydrogen atom in quantum mechanics—upon which more realistic scenarios involving radiation can be built, potentially using perturbation theory.

The hydrogen atom serves as more than a historical benchmark; its exact analytical solution and high-precision spectrum are indispensable for ongoing tests of fundamental constants~\cite{Webb:2000mn,Kanekar:2018mxs} and quantum principles. Our cosmological model, by virtue of its precise mapping to this keystone system, is endowed with this same analytical potency. Just as atomic hydrogen shaped quantum theory, this framework provides a \emph{precision tool} for quantum gravity, with implications ranging from AdS holography to singularity resolution.  
With the analytical power of hydrogen now brought to bear on the cosmos, the stage is set for significant advances in quantizing spacetime.

\section*{Acknowledgments}
HSS and DM thank Kinjalk Lochan for his helpful comments and suggestions. 
HSS is thankful to Suprit Singh for helpful comments. DM is thankful to Shiv K.~Sethi for interesting discussions. The research of HSS is supported by the Core Research Grant CRG/2021/003053 from the Science and Engineering Research Board (SERB), India. DM thanks Raman Research Institute for support through postdoctoral fellowship. DM's and HSS's visits to IIT Bombay was supported by SERB-CRG grant CRG/2022/002348 of SS. 

\bibliographystyle{utphys}
\bibliography{bib_resource.bib}

\end{document}